# Cathodoluminescence as an Effective Probe of Carrier Transport and Deep Level Defects in Droop-Mitigating InGaN/GaN Quantum Well Heterostructures

Zhibo Zhao[1,a)], Akshay Singh[1,a)], Jordan Chesin[1], Rob Armitage[2], Isaac Wildeson[2], Parijat Deb[2], Andrew Armstrong[3], Kim Kisslinger[4], Eric A. Stach[4,b)], Silvija Gradečak[1,c)]

[1]Department of Materials Science and Engineering, Massachusetts Institute of Technology, Cambridge, Massachusetts, 02139, USA

[2]Lumileds LLC, San Jose, California, 95131, USA

[3]Sandia National Laboratories, Albuquerque, New Mexico, 87185, USA

[4]Center for Functional Nanomaterials, Brookhaven National Laboratory, Upton, New York, 11973, USA

Commercial InGaN/GaN light emitting diode heterostructures continue to suffer from efficiency droop at high current densities. Droop mitigation strategies target Auger recombination and typically require structural and/or compositional changes within the multi-quantum well active region. However, these modifications are often accompanied by a corresponding degradation in material quality that decreases the expected gains in high-current external quantum efficiency. We study origins of these efficiency losses by correlating chip-level quantum efficiency measurements with structural and optical properties obtained using a combination of electron microscopy tools. The drop in quantum efficiency is not found to be correlated with quantum well (QW) width fluctuations. Rather, we show direct correlation between active region design, deep level defects, and delayed electron beam induced cathodoluminescence (CL) with

---
a) These authors contributed equally to this work.
b) Present address: Department of Materials Science and Engineering, University of Pennsylvania, Philadelphia, PA 19104, USA
c) Author to whom correspondence should be addressed. Electronic mail: gradecak@mit.edu

characteristic rise time constants on the order of tens of seconds. We propose a model in which the electron beam fills deep level defect states and simultaneously drives reduction of the built-in field within the multi-quantum well active region, resulting in a delay in accumulation of carrier populations within the QWs. The CL measurements yield fundamental insights into carrier transport phenomena, efficiency-reducing defects, and quantum well band structure that are important in guiding future heterostructure process development.

InGaN/GaN quantum well (QW) heterostructures have enabled inorganic light emitting diodes (LEDs) with external quantum efficiencies (EQE) exceeding 70%.[1,2] However, InGaN/GaN LEDs typically achieve peak EQE at relatively low current densities of ~ 10 A/cm$^2$. At higher drive currents, the EQE decreases monotonically in a phenomenon known as efficiency droop.[3] Overcoming droop in the high current regime (>100 A/cm$^2$) can yield significant gains in both cost and wall-plug efficiency, particularly in emerging high power applications.

A number of mechanisms have been proposed to account for (non-thermal) efficiency droop including current crowding, carrier delocalization, electron overshoot, and Auger recombination.[3-5] Parallel droop mitigation strategies typically require structural and/or compositional changes within the active region QWs and barriers to improve carrier spreading and enhance the radiative recombination rate. However, these modifications are often accompanied by a pronounced reduction in low-current EQE suggesting increased Shockley-Read-Hall (SRH) defect densities.[6,7] To realize actual gains in EQE, droop mitigating design concepts must be implemented without significant material quality degradation. The optoelectronic nature of defects in InGaN alloys, including deep level charge traps important in SRH pathways[6,7] and V-pits present within the multi-QW region[8-10], remains controversial and continues to hinder development of next-generation high-power LED devices and phosphor-free lighting solutions. Similar defects dominate non-radiative recombination in indium-rich green InGaN LEDs, which are necessary for bridging the green gap in red-green-blue lighting applications.[11-14]

Electron microscopy,[15-17] in conjunction with standard device-level electrical characterization, allows direct correlation between QW microstructure, optical emission, and chip-scale EQE. In this paper, we use cathodoluminescence in a scanning electron microscope (SEM-CL) to measure optical properties of LEDs comprised of InGaN/GaN QW heterostructures designed to reduce Auger recombination and mitigate efficiency droop. A time-delayed response (tens of seconds) from the InGaN QWs was observed for a subset of samples, while luminescence from the underlying GaN and dilute InGaN layers appeared instantaneously (within ~ 1 s). We find that neither the delayed CL response nor chip-scale EQEs are directly governed by microscopic structural modulations (specifically, QW width fluctuations). Instead, the delayed CL dynamics are attributed to a combination of (i) carrier sweep-out and subsequent charge accumulation across active region and (ii) carrier capture by deep level traps in vicinity of the QW active region. Hence, the observed time-delayed SEM-CL yields qualitative insights into device-relevant carrier transport within the multi-QW layers.

Two series of low-droop InGaN/GaN QW heterostructures emitting at ~ 450 nm were grown epitaxially via metal-organic chemical vapor deposition (MOCVD) on sapphire substrates. Simplified schematics of the LED epitaxial structures are shown in Figure 1a. The first series of LEDs, HP10 and LP10, consists of a ten-period multi-QW active region with 3 nm InGaN QWs sandwiched between GaN barriers. In order to study effects of extended structural inhomogeneities[14,16] on optoelectronic properties, QW width fluctuations (localized changes in QW thickness) were deliberately introduced into these samples. LP10 included changes to epitaxy process that were expected to result in different droop and SRH recombination behavior compared to HP10. Representative EQE measurements (Figure 1b) demonstrate that the device LP10 (LP, low performance) has significantly lower EQE compared to HP10 (HP, high performance).

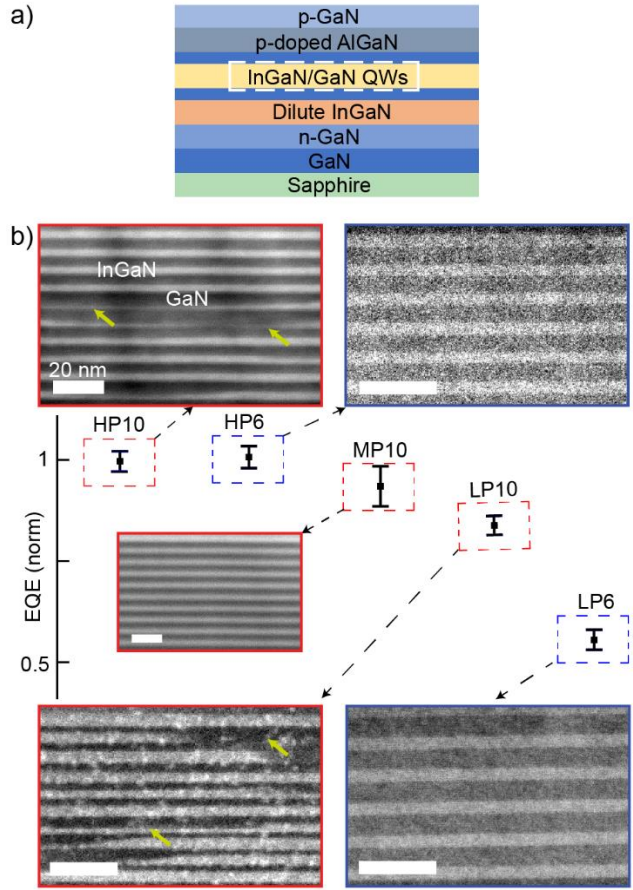

**Figure 1.** a) Schematic of different layers in LED samples. b) EQE values (normalized to EQE of HP10), and corresponding representative dark field STEM images of InGaN/GaN QW regions (dashed box in (a)) for HP10, HP6, MP10 (shown as inset), LP10, and LP6. QW width fluctuations are indicated by yellow arrows in STEM images. All scale bars are 20 nm.

We next performed scanning transmission electron microscopy (STEM) characterization of the device active regions in correlation with their macroscale efficiency. Sample preparation details are provided in the supplementary material. STEM was performed using either an FEI Titan or Hitachi 2700C, operating at 80 kV or 120 kV respectively. Importantly, the accelerating voltage was maintained below knock-on damage threshold for InGaN alloys to prevent beam-induced structural changes.[18] Since mass-thickness is the dominant contrast mechanism in dark field STEM, InGaN QWs appear brighter compared to GaN barriers, and representative STEM images confirm the presence of QW width fluctuations in HP/LP10 (Figure 1b). Although these STEM images show limited areas of the devices, our conclusions are based on a statistically significant number of images obtained across larger areas.

While these samples have same number of QWs and similar QW/barrier thicknesses, the presence of QW width fluctuations does not correlate with chip-scale EQE measurements. In order to highlight the disparity between gross-scale structural imperfections and EQE, a third LED heterostructure (MP10) devoid of QW width fluctuations was grown in the same series as HP/LP10. The EQE of MP10 (MP, medium performance) was measured to be intermediate between HP10 and LP10 despite no noticeable QW width fluctuations.

To further investigate the possible role of QW period and barrier thickness, we prepared a second series, HP6 and LP6, consisting of six-period multi-QW active region with 3 nm InGaN QWs sandwiched between 5 nm GaN barriers. Because QW width fluctuations were found to be unrelated to EQE, no effort was made to intentionally introduce fluctuations for HP/LP6 (Figure 1b). Similar to the previous comparison, a change in epitaxy processes of LP6 and HP6 was made with the expectation that the two samples would show different droop and SRH recombination behaviors. As expected, EQE measurements confirm that LP6 devices have significantly lower EQE compared to HP6, while HP6 exhibited the highest median EQE of all samples (Figure 1b). Taken together, chip-scale EQE and electronic carrier properties within QW active region are not related to features that are either a function of QW barrier thickness/period or immediately obvious from STEM-based structural analysis.

In order to directly probe optoelectronic carrier properties within the LED heterostructure, we next performed SEM-CL measurements. Briefly, an electron beam is incident on the wafer along the growth direction, generating electron-hole pairs that subsequently recombine radiatively and produce CL. The electron penetration depth can be changed by tuning accelerating voltage, thus providing control over relative contributions from different layers of the heterostructure.[19,20] SEM-CL was measured in a JEOL JXA-8200 Superprobe operated at an accelerating voltage of 10 kV with beam current of 1 nA, unless specified otherwise. SEM-CL spectra were acquired once per second using 100 ms integration time under continuous spot mode irradiation (of a single spot). Multiple spot measurements were obtained for each device to ensure results were representative of the entire device.

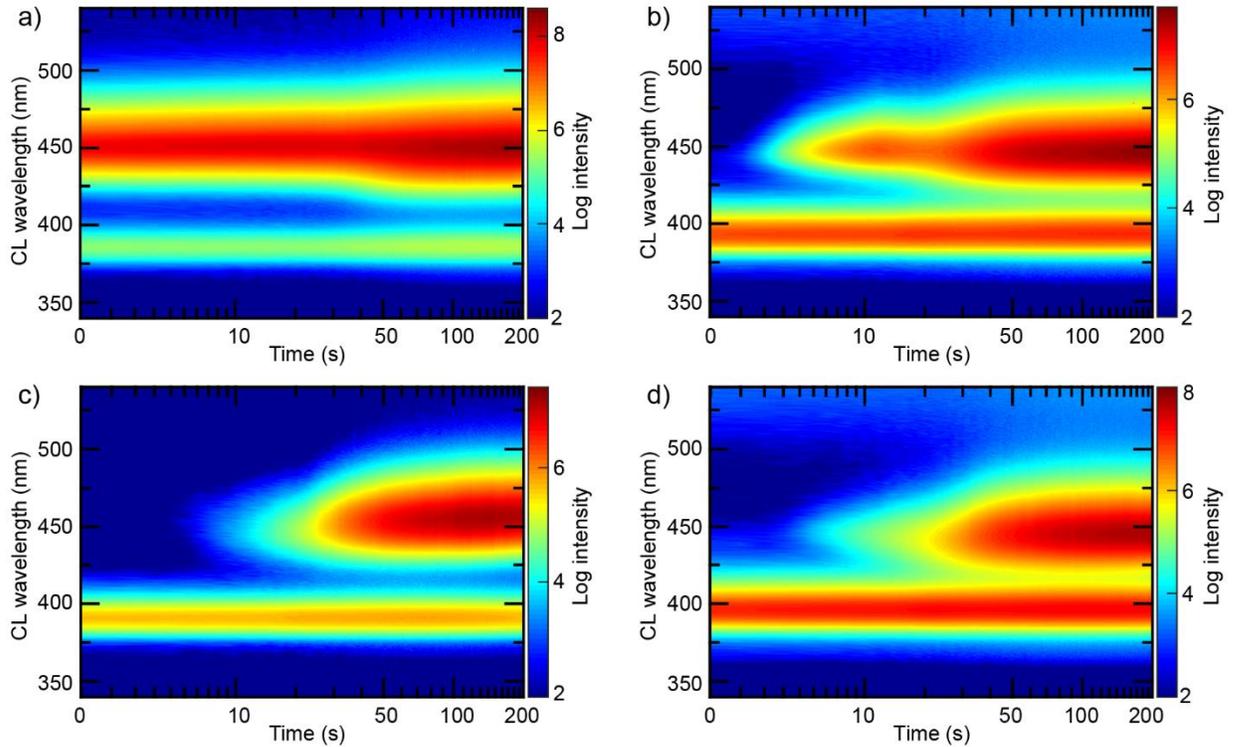

**Figure 2.** a)-d) Time dependent CL for HP10, HP6, LP10, and LP6, respectively. Spectral region corresponding to dilute InGaN underlayer (395 nm) and QWs (~ 450 nm) is shown. Time is plotted on log scale to elucidate initial rise dynamics.

CL spectra for HP10, HP6, LP10, and LP6 as a function of electron beam irradiation time are shown in Figures 2(a-d) respectively. The peak emission from the QW region is centered at ~

450 nm for all samples. A weak GaN peak centered at 365 nm and dilute InGaN underlayer peak centered at 395 nm are also observed. The weak yellow band luminescence (not shown) is observed as a broad peak centered at 550 nm.[21-23] A representative spectrum is shown in Figure S1 in the supplementary material.

For HP6, LP6, and LP10, the initial intensity of QW emission is zero whereas the dilute InGaN signal appears instantaneously. The QW emission continues to rise upon further electron beam irradiation and saturates after ~ 120 s. In contrast, intensity of QW emission in HP10 starts instantaneously at 60% of its maximum value and saturates over the course of measurement.

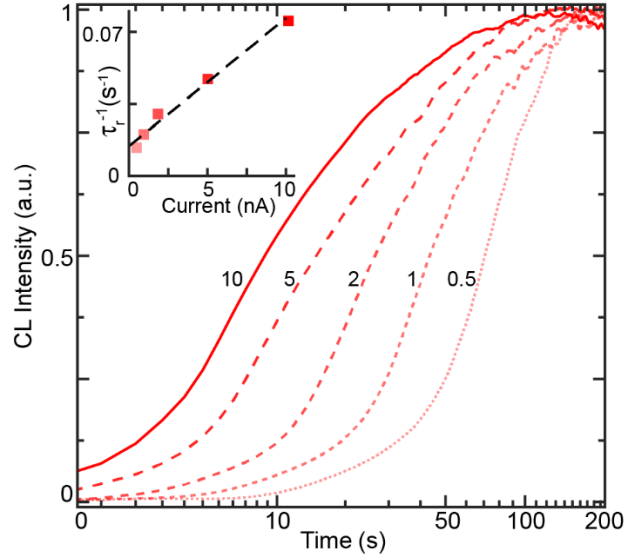

**Figure 3**. QW rise time response for LP6 as a function of beam current. The beam currents (nA) are indicated on each curve. The curves are normalized to maximum CL for each current. The inset shows rise rates ($\tau_r^{-1}$) as a function of beam current.

Representative current-dependent SEM-CL measurements on LP6 reveal a strong dependence of CL rise time on incident probe current (HP6 and LP10 exhibit similar rise dynamics). LP6 was exposed to 0.5, 1, 2, 5, and 10 nA probe currents, and intensity of QW emission was monitored over the course of 200 s. The rise in QW emission intensity follows a distinct S-shaped curve, starting from zero for probe currents of 2 nA or less, and small finite values at higher probe currents. The CL rise can be described by an empirically motivated logistic function:

$$I_{CL} = \frac{I_0}{1 + \exp[-k(t - \tau_r)]} \tag{1}$$

where $\tau_r$ is defined as rise time constant (tens of seconds), with $I_0$ and $k$ as fitting parameters (Table S2). Plotting rise rate ($\tau_r^{-1}$) as a function of beam current shows a positive linear correlation (Figure 3, inset) indicative of carrier injection-governed behavior.[24,25]

The observed CL characteristics suggest electron beam-induced changes in carrier dynamics in vicinity of the QW active region. These effects are likely confined to the QW active region because only the QW emission intensity changes under electron beam irradiation, whereas the GaN peak, dilute InGaN underlayer, and yellow band emission intensities remain constant. Furthermore, electron acceleration voltage-dependent measurements show that CL rise dynamics are faster for an accelerating voltage of 10 kV ($\tau_r$ =48.9±0.4 s) compared to 25 kV ($\tau_r$ =88.2±0.1 s), attributable to variable degrees of interaction between the electron beam and QW active region at different voltages (Figure 4a). Indeed, Monte Carlo simulations[26] of beam-sample interactions confirm a stronger interaction between beam and QW active region at 10 kV, yielding faster rise dynamics (Figure 4b). Quantitatively, Monte Carlo simulations estimate that 15% of total CL intensity is generated within QW region at 10 kV, as opposed to only 2% at 25 kV.

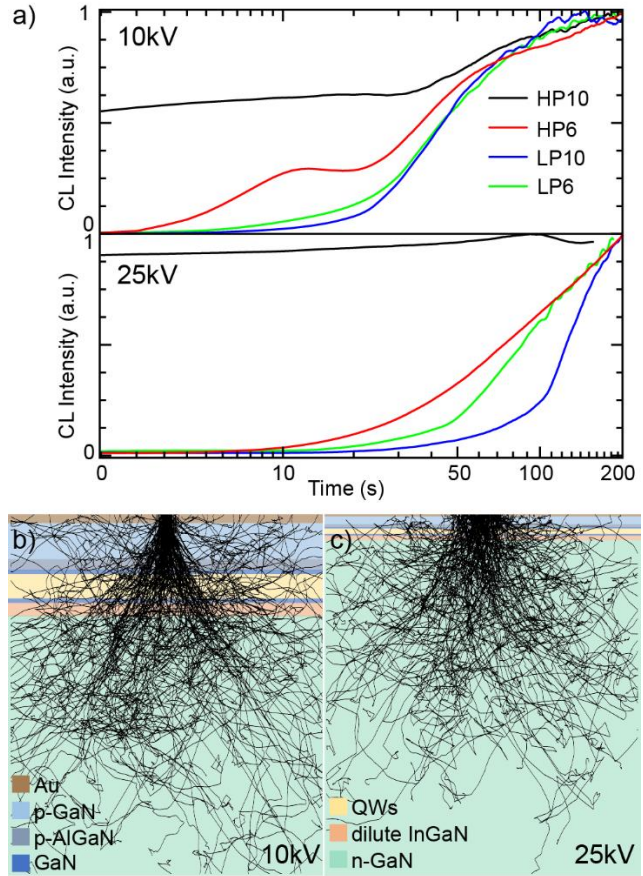

**Figure 4.** a) Electron beam voltage-dependent CL dynamics. All signals are normalized to their respective CL maxima. With increasing voltage, longer rise times are observed for CL saturation. b) Monte Carlo simulations of incident electron scattering paths, overlaid with LED structure. A thin (~ 20 nm) gold layer is deposited on top to prevent charging in CL measurements.

We attribute the delayed CL response to a combination of progressive screening of built-in field and deep level trap filling. Since the QW active region is embedded between heavily doped GaN layers, the QW region resembles a p-i-n junction. When the electron beam is initially incident on the sample, electron-hole pairs generated inside QW active region are swept apart by the built-in field, preventing radiative recombination. These separated carriers produce a time-dependent counter-voltage that screens the built-in field. As carriers are continuously generated under the electron beam, the screening effect intensifies until the junction reaches a steady-state open-circuit voltage. Thus, carriers generated at later times are more likely to undergo radiative recombination within QW region due to stronger screening of built-in field. Consequently, QW emission intensity increases over time until steady-state is reached.

The evidence for charge separation and accumulation is twofold. First, we fabricated a sample structurally identical to LP10, but without p-doped layers. Because this sample lacks a p-n junction, generated carriers do not experience a strong built-in field and are more likely to undergo radiative recombination from the onset of CL excitation. Indeed, QW emission of this junction-free sample did not exhibit delayed saturation dynamics of the full device (Figure S3). Second, strength of the built-in field is smaller for devices with larger total thickness of (intrinsic) active region, which should result in faster saturation of CL. The built-in field in HP10, for example, is expected to be at least 35% smaller than for any other sample due to its higher (intrinsic) active region thickness. The absence of delayed saturation dynamics in HP10 corroborates the idea that carriers generated within active region of this device are less susceptible to built-in field induced sweep-out.

Remarkably, we note that the QW CL enhancement (after electron irradiation) observed in LP6, HP6, and LP10 remains irreversible for at least 30 min (Figure S5). These observations can be attributed to carrier capture by long-lived deep level traps near the QW active region. Under the proposed model, carriers captured by deep level traps within the surrounding doped GaN layers sustain a non-equilibrium charge carrier density on either side of the junction. Since sustained radiative recombination requires both trap filling and charge screening, the two processes mutually contribute to significant delay in the CL. We speculate that such deep level traps might share a similar nature with metastable defects in GaN associated with persistent changes in photoconductivity.[27-30] These metastable defects have a strong electron-phonon coupling such that changes in the defect charge state are coupled with lattice relaxations whose effects can persist for hours to days. These defects may originate from highly non-equilibrium growth conditions for the heterostructures of interest.[31]

The final feature of interest in SEM-CL spectra is the initial fast rise (~ 10 s) followed by a brief plateau in QW emission observed only in HP6 and HP10 (Figure 4a). We attribute this feature to a pre-existing population of carriers in one or more QWs in the active region of HP6 and HP10 under open circuit condition. To confirm this, capacitance-voltage (C-V) measurements were performed under reverse bias on HP6, LP6, HP10, and LP10 (Figure S6). The C-V data suggest that the Fermi level of one or more QWs in HP6 and HP10 lies near or within the conduction band, resulting in significant open-circuit carrier density inside the QW region. In pre-populated QWs, carriers excited by electron beam can recombine with pre-existing carriers, so the initial rise and plateau reflect rapid saturation of radiative recombination within these pre-populated QWs followed by slower saturation across remaining QWs.

In the current work, we have measured CL dynamics for a series of droop minimizing InGaN LED heterostructures. We observe that EQE and electronic carrier properties are not directly governed by microscopic structural imperfections. Instead, dynamical features of the SEM-CL spectra can yield device-relevant insights into (i) carrier escape and confinement within the QWs; (ii) deep level trap states; and (iii) QW band structure. First, the delayed CL reflects drift of carriers out of the QWs. While drift should significantly depend on both QW/barrier design and built-in electric field strength across active region, in comparisons of samples where the electric field is a controlled variable the CL rise behavior can be studied to infer differences in carrier transport between different active region designs. While enhanced confinement within QWs improves the radiative recombination rate and EQE in low to intermediate current regimes, carrier spreading is desirable at high current densities to reduce carrier density per QW and mitigate droop. Second, delayed saturation dynamics are attributed to carrier capture by deep level traps. Filling these long-lived traps manifests as a persistent effect which could possibly impact steady-state device operation. In-depth device simulations may elucidate and quantify the effects of deep level traps on CL rise time constant. Finally, certain features in SEM-CL spectra are reflective of the carrier population within one or more QWs, providing clues to band structure and Fermi level position in specific QWs under open circuit conditions. Ultimately, insights gained through SEM-CL can help identify mechanisms limiting device performance and guide heterostructure growth and process development in more detail than standard EQE measurements.

**Supplementary material**: details on sample preparation and imaging, comprehensive fit parameters for different probe currents, details of Monte Carlo simulations, CL rise dynamics of the junction-free sample, schematic of the proposed carrier transport model, recovery dynamics, and plotted C-V data.

**Acknowledgements**: This material is based upon work supported by the Department of Energy, Office of Energy Efficiency and Renewable Energy (EERE), under Award Number DE-EE0007136. Z. Zhao also acknowledges support through a National Science Foundation Graduate Research Fellowship under Grant Number 1122374. This research used resources of the Center for Functional Nanomaterials, which is a U.S. DOE Office of Science Facility, at Brookhaven National Laboratory under Contract No. DE-SC0012704. Sandia National Laboratories is a multi-mission laboratory managed and operated by National Technology and Engineering Solutions of Sandia, L.L.C., a wholly owned subsidiary of Honeywell International, Inc., for the U.S. Department of Energy's National Nuclear Security Administration under contract DE-NA-0003525. This work made use of the Shared Experimental Facilities supported in part by the MRSEC program of the National Science Foundation under award number DMR-1419807. The authors thank Dr. N. Chatterjee for assistance with SEM-CL measurements.**References**

[1] C. J. Humphreys, MRS Bulletin **33**, 459 (2008).

[2] N. Yukio, I. Masatsugu, S. Daisuke, S. Masahiko, and M. Takashi, J. Phys. D: App. Phys. **43**, 354002 (2010).

[3] J. Piprek, Physica Status Solidi (a) **207**, 2217 (2010).

[4] S. Karpov, Opt. and Quantum Electronics **47**, 1293 (2015).

[5] J. Iveland, L. Martinelli, J. Peretti, J. S. Speck, and C. Weisbuch, Phys. Rev. Lett. **110**, 177406 (2013).

[6] D.-P. Han, D.-G. Zheng, C.-H. Oh, H. Kim, J.-I. Shim, D.-S. Shin, and K.-S. Kim, Appl. Phys. Lett. **104**, 151108 (2014).

[7] A. Armstrong, T. A. Henry, D. D. Koleske, M. H. Crawford, and S. R. Lee, Opt. Exp. **20**, 812 (2012).

[8] M. Kim, S. Choi, J. H. Lee, C. Park, T. H. Chung, J. H. Baek, and Y. H. Cho, Scientific Rep. **7**, 42221 (2017).

[9] C.-Y. Chang, H. Li, Y.-T. Shih, and T.-C. Lu, Appl. Phys. Lett. **106**, 091104 (2015).

[10] D. I. Florescu, S. M. Ting, J. C. Ramer, D. S. Lee, V. N. Merai, A. Parkeh, D. Lu, E. A. Armour, and L. Chernyak, Appl. Phys. Lett. **83**, 33 (2003).

[11] C. J. Humphreys, J. T. Griffiths, F. Tang, F. Oehler, S. D. Findlay, C. Zheng, J. Etheridge, T. L. Martin, P. A. J. Bagot, M. P. Moody, D. Sutherland, P. Dawson, S. Schulz, S. Zhang, W. Y. Fu, T. Zhu, M. J. Kappers, and R. A. Oliver, Ultramicroscopy **176**, 93 (2017).

[12] K. A. Bulashevich, A. V. Kulik, and S. Y. Karpov, Physica Status Solidi (a) **212**, 914 (2015).
14

# Supplementary

**I. Sample Preparation**

For high resolution measurements cross-sectional scanning transmission electron microscopy (STEM) imaging, thin (< 100 nm) lamellae of light emitting diode (LED) samples were prepared using a focused gallium ion beam in a Helios Nanolab system. The lamellae were thinned with successive reductions in ion beam accelerating voltages (30 kV, 16 kV, 5 kV, 2 kV) to minimize beam damage and ion implantation. The final step at 2 kV removes surface amorphization.

As illustrated in the main text (Figure 1b), we observed quantum well (QW) width fluctuations in LEDs HP10 and LP10, but not in HP6 and LP6. For comparison of structure across these four samples, we have assigned a structural integrity index to these samples (summarized in Table S1), with 0 representing severe QW width fluctuations, and 2 representing no fluctuations. The index is measured by quantifying QW width fluctuations for a number of lamellae to yield statistically significant conclusions. We observe that LP10 and LP6/HP6 have the lowest and highest structural integrity, respectively.

**Table S1**: Summary of the LED samples under study. The structural details are indicated, as well as spectral peaks for QWs.

| Sample | # of QW | QW peak wavelength | Structural Integrity |
|---|---|---|---|
| HP10 | 10 | 448 nm | 1 |
| HP6 | 6 | 447 nm | 2 |
| LP10 | 10 | 455 nm | 0 |
| LP6 | 6 | 444.6 nm | 2 |

**II. SEM-CL Methods**

Cathodoluminescence in a scanning electron microscope (SEM-CL) was measured in a JEOL JXA-8200 Superprobe operated at accelerating voltages of 10-25 kV. The beam current was varied between 0.5-10 nA to study the effect of charge injection rate on CL rise dynamics. The cathodoluminescence was collected by a parabolic mirror situated just above the sample. Each LED sample was exposed to a stationary electron beam over the course of 200 s, and SEM-CL spectra were acquired once per second (with 100 ms integration time) under continuous spot mode illumination (100 nm probe diameter). In Figure S1, we illustrate a typical spectrum for LP6 (at time ~ 100 s), with four peaks as discussed in the main text.

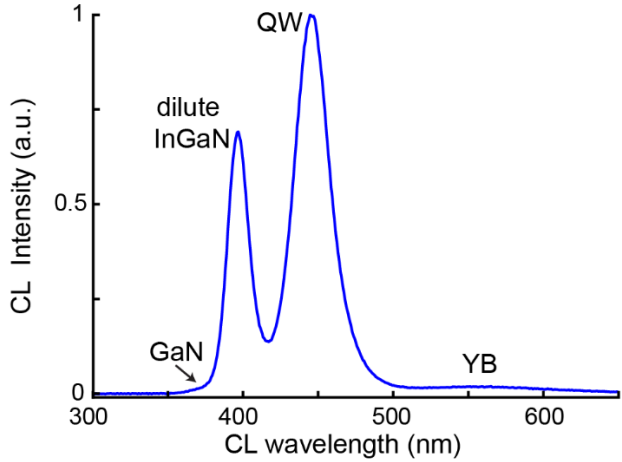

**Figure S1.** A typical spectrum for LP6, illustrating the GaN (365nm), dilute InGaN (395nm), QW (450nm) and the broad yellow band (YB, 500-600 nm) luminescence peaks.

### III. Fit Parameters for Varying Probe Currents

The fitting function used for extracting time constants for the QW rise dynamics is a logistic function of the form

$$I_{CL} = \frac{I_0}{1 + \exp[-k(t - \tau_r)]}$$

where $\tau_r$ is defined as the rise time (time taken for half amplitude), and $I_0$ (amplitude ~ 1), and $k$ (slope) are fitting parameters. This function has been used extensively for modelling rate of growth in material science and population modelling in biological experiments.

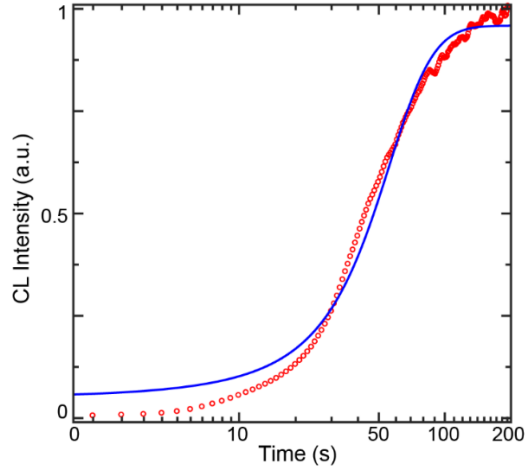

**Figure S2**: Red markers indicate the QW rise time response for LP6 at current =1 nA. Blue line indicates the logistic fit to the data.

In Figure S2, the QW rise time response (red markers) for LP6 at a current value of 1 nA is illustrated, along with the blue logistic fit curve. Further, fitting parameters for different currents for LP6 are indicated in Table S2. The amplitude ($I_0$) is close to 1 for all cases (consequence of normalization to maximum amplitude). The slope ($k$) increases with increasing currents, and is related to variations in shape of the curve. A consistent decrease of $\tau_r$ as a function of the beam current is observed, as indicated in the main text.

**Table S2**: Fitting parameters for LP6 QW rise time response for different currents.

| Current (nA) | $I_0$ (a.u.) | $\tau_r$ (s) | $k$ (1/s) |
|---|---|---|---|
| 0.5 | 0.99±0.01 | 75.2±0.2 | 0.048±0.001 |
| 1 | 0.96±0.01 | 48.9±0.4 | 0.059±0.001 |
| 2 | 0.95±0.01 | 32.3±0.4 | 0.073±0.002 |
| 5 | 0.96±0.01 | 21.1±0.5 | 0.075±0.003 |
| 10 | 0.97±0.01 | 13.3±0.4 | 0.106±0.004 |

## IV. Monte Carlo Simulations

The Monte Carlo simulations were performed using the Casino 2.48 software package [1]. The thicknesses of different layers of the heterostructure and elemental compositions were input as starting parameters, and a large number of incident electrons ($N \sim 10,000$) were simulated. The relevant parameters that can be extracted as a function of accelerating voltage are the contribution of different layers towards CL and range of incident electrons. The contribution of the QW region towards CL is maximum for 10 kV, and decreases for other voltages, in agreement with our observations. As noted in the main text, Monte Carlo simulations estimate that 15% of the total CL intensity is generated within the QW region at 10 kV, as opposed to only 2% at 25 kV.

## V. Effect of Built-in Field

To study the effect of built-in field on the CL dynamics, we fabricated a sample structurally identical to LP10, but without p-doped layers above the QW region. This sample is thus grown without a p-n junction, and consequently has a much reduced built-in field. As a result, the electron beam-generated electron-hole pairs are not swept apart, which increases the probability of radiative recombination. Indeed, the QW emission of this junction-free sample was observed to be stable over time (Figure S3) and did not exhibit the delayed saturation kinetics of the full device.

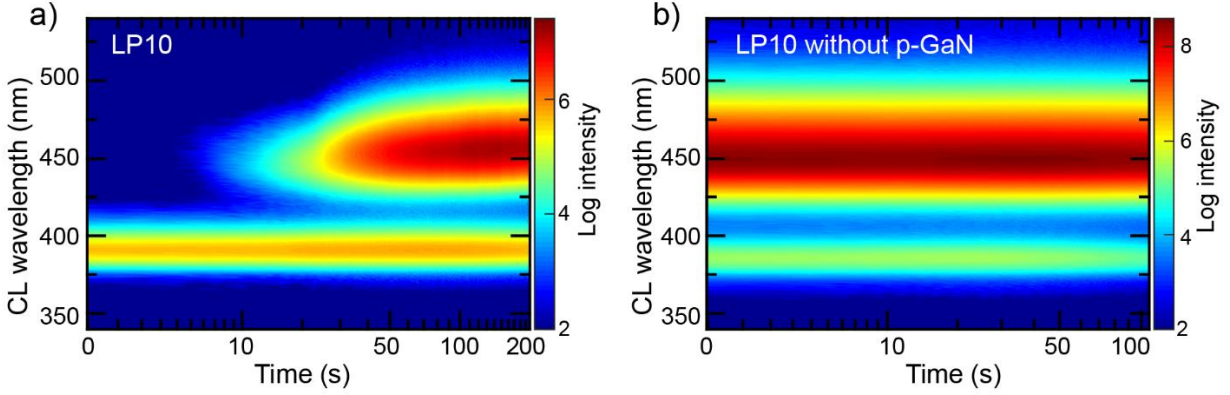

Figure S3: Time dependence of CL for LP10 a) full device, b) without the top p-GaN. The time axis for b) is shorter (100 s c.f. 200 s for (a)) since no rise time dynamics were observed. As in the main text, time is plotted on a log scale.

## VI. Schematic of Proposed Carrier Model

In this section we present a schematic (Figure S4) for the model explained in the main text. We attribute the delayed CL response to a combination of progressive screening of the built-in field and deep level defect passivation. Under the proposed model, carriers captured by deep level traps in the vicinity of QW active region are not swept out, and thus do not significantly contribute to screening of the built-in field. Similarly, carriers swept out by the built-in field are not captured by deep level traps and do not contribute to defect passivation. Since radiative recombination requires both charge screening and defect passivation, the two processes mutually contribute to the significant delay in the CL response.

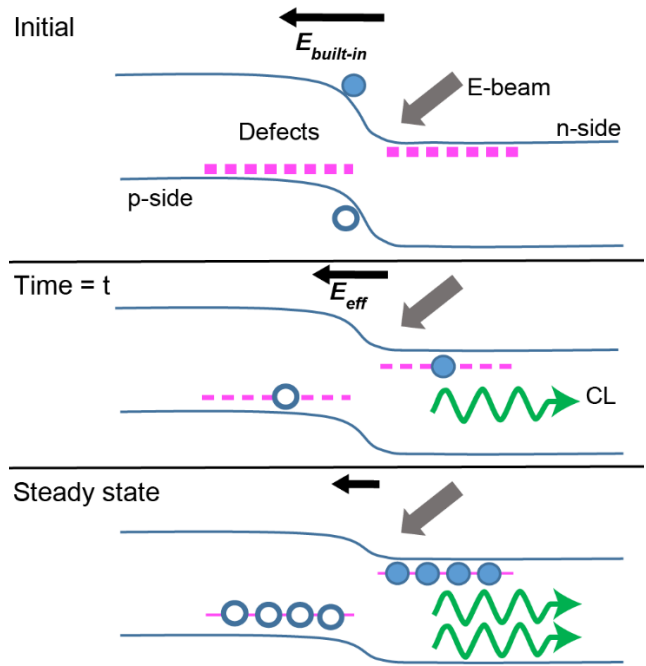

**Figure S4.** Illustration depicting dynamics of continuous electron beam irradiation on counter-voltage and deep level defect passivation. The thickness of the defect line (in the figure) represents the number of unoccupied defect level states. Consequently, CL increases with time until steady state is reached.

## VII. Recovery Dynamics

To study whether the effect of electron beam persists even after the beam is blocked, we measured the recovery behavior of CL on LP6 (LP10 and HP6 show similar behavior). After the initial 200 s electron beam exposure (as described in the main text), the sample was allowed to recover by blocking the electron beam for 3 min. CL was then measured on the same spot for 30 s with time steps of 5 s, with the electron beam blocked in between spectra collection steps. Note that no CL is detected when the electron beam is blocked. A longer gap of 3 min was also used, wherein the beam is blocked again. This measurement cycle- consisting of 3 min beam block, spectra taken for 30 s with 5s time step, 3 min beam block was repeated. Thus, we used a combination of 5 s and 3 min beam blocking time steps to study short-time and long-time recovery behavior. As seen in Figure S5, the CL signal does not recover to its initial value, on either a short time scale or even after ~ 30 min. The lack of recovery on observable time scales is an indication of metastable long-lived trap states, as discussed in the main text.

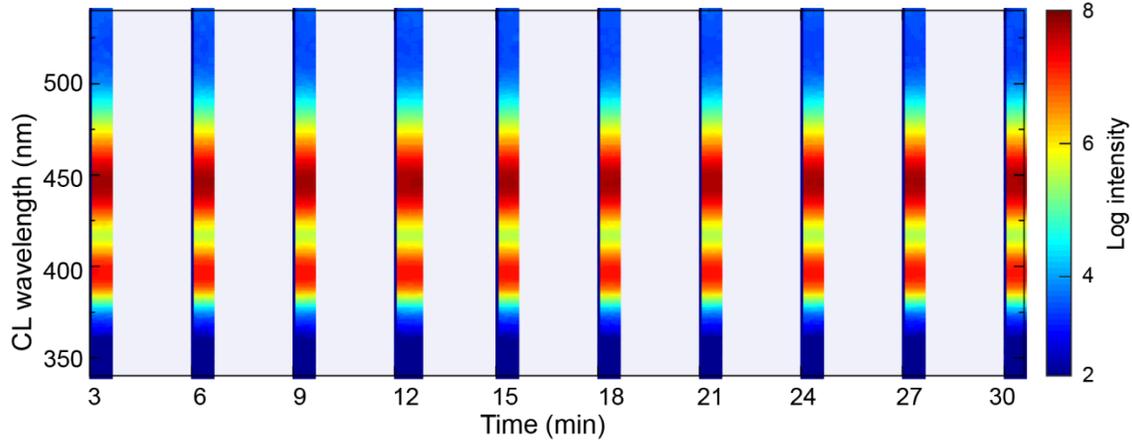

**Figure S5**. CL of LP6 plotted after initial 200 s exposure. As in Figure S3, the log of intensity is plotted. Time is plotted on a linear scale. Spectra are measured for 30 s with time steps of 5 s, and also with longer time gaps of 3 min. In between (light shaded area), the electron beam is blocked and sample is allowed to recover.

## VIII. Capacitance-Voltage (C-V) data

Capacitance-voltage (C-V) measurements were performed under reverse bias on all LEDs of interest. These measurements allow investigation of charge carrier density in different layers of the heterostructure device. In Figure S6, we illustrate CV measurements on LP6 and HP6. The lightly doped QW region is immediately apparent as it has lower charge density than the heavily doped surrounding n- and p-GaN layers. A significant charge carrier density within the QW region is observed in HP6 (and HP10), but not in LP6 (and LP10).

The C-V data suggest that the Fermi level of one or more QWs in HP6 (and HP10) lies near or within the conduction band. The initial fast rise (around ~ 10 s) for HP6 (and HP10) (Figure 4a), is thus attributed to a pre-existing population of carriers in one or more QWs in the active region. QWs prepopulated with carriers are predisposed to faster radiative recombination. Thus the initial rise and plateau demonstrate saturation of radiative recombination within these prepopulated QWs, followed by slower saturation across the remaining QWs.

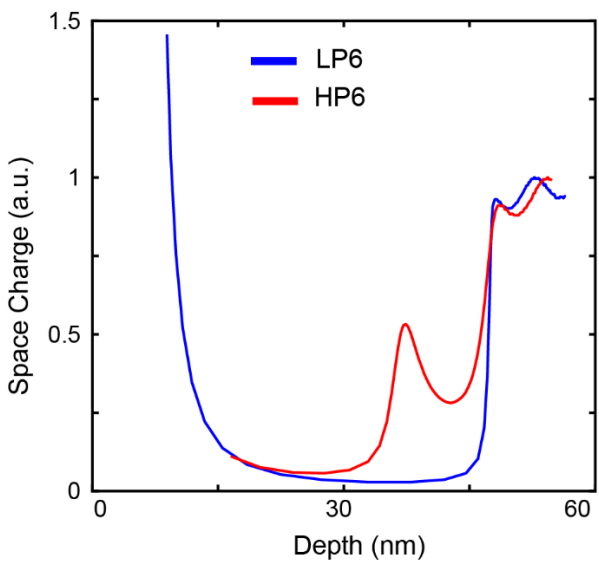

**Figure S6**. Capacitance-voltage data for LEDs LP6 and HP6. The charge density (space charge) is plotted as a function of distance along the growth direction of the LED.